\documentstyle[12pt]{article}
\def\be{\begin{equation}}
\def\ee{\end{equation}}
\begin{document}
\title{
\begin{flushright}
{\small SMI-08-98 }
\end{flushright}
\vspace{2cm}
Large $N$  Gauge Theories\\ and\\
Anti-de Sitter Bag Model}
\author{
I.V. Volovich
\thanks{Permanent address: Steklov Mathematical Institute,
 Gubkin St.8, GSP-1, 117966, Moscow, Russia, volovich@mi.ras.ru}
\\
$~~~~$  \\
{\it Centro Vito Volterra}\\
{\it Universita di Roma Tor Vergata, Rome, Italy}}

\date {$~$}
\maketitle
\begin {abstract}
Using a proposal of Maldacena one describes the large $N$ limit
of gauge theories in terms of supergravity solutions on anti-de Sitter
space. From this point of view we discuss a  possible scenario for
quark confinement in gauge theory by
describing hadrons as strongly curved universes.
In particular an interpretation of black hole as a bag model
in SQCD is discussed. One relates the mystery of curvature singularities
in classical general relativity with the mystery of quark confinement.
The {\it AdS} bag model is defined by computing the probe membrane
action in supergravity background. It naturally implies the
"Cheshire Cat bag" principle.
The confining pressure in the MIT bag model is
related with the cosmological constant in the {\it AdS} bag model.
The Skyrme model is interpreted as an effective theory describing
black holes.
\end{abstract}

\newpage
In many recent works a relation between the large $N$ limit
of supersymmetric Yang-Mills theories and supergravity
has been explored \cite{Mal}-\cite{Oz}. It was proposed by Maldacena
\cite{Mal} that in particular the large $N$ limit of the ${\cal N}=4$
super Yang-Mills theory in four dimensions with $U(N)$ gauge
group is equivalent to the type IIB string theory on $AdS_5\times S^5$.
The gauge theory describes a region of spacetime near the horizon
of a black hole which is known to have anti-de Sitter $(AdS)$ geometry.
Moreover Horowitz and Ooguri \cite{HO} have argued that the gauge theory
actually describes string theory on both sides of the horizon.
Therefore one has to find in gauge theory an analogue of the
horizon which is the basic notion of the black hole physics.

In this note we discuss a scenario for a very natural
possible correspondence  between basic notions of gauge theory
and gravity. It is tempting to speculate that  confinement in gauge
theory corresponds to black holes in gravity and
one can try to describe hadrons as strongly curved universes
or black holes. In this picture the horizon
corresponds to the surface of the bag in a bag model.
One relates the mystery of curvature singularities
in classical general relativity with the mystery of quark confinement.

It is proposed in \cite{HR} that the gauge theory could provide
a resolution of the singularity inside the black hole since
it is perfectly regular. From our point of view the gauge theory is not
a regular theory, rather it is a very hard theory
especially in the infra-red regime.
One believes that perhaps the gauge theory could really give some help
for understanding the curvature singularities in general relativity
but  only if we are able to understand the problem of confinement.

The bag model was one of the first attempts to understand the color
confinement \cite{Bog,MIT}. In bag models quarks and gluons are
confined to a bounded region that usually is taken as a static sphere
of radius $R$.
In the MIT bag model \cite{MIT}
the bag exterior acts as a perfect color dielectric, while
the color degrees of freedom are confined to the bag interior.
At the surface of the bag, the gluons are subject to the confining
boundary conditions
\be
\label{1}
n E^a=0, ~~n\times B^a=0
\ee
where $E^a$ and $B^a$ are color electric and magnetic fields
and $n$ is the outward normal to the bag surface.
For the quarks, the confining boundary condition states
that there is no flux of the isoscalar current through the
bag surface:
\be
\label{2}
n_i{\bar q}\gamma^i q=0
\ee
The bag model gives a rather satisfactory description of masses,
magnetic momenta and some other properties
of hadrons. One of the main problems with the bag model is that
it is unclear how to derive the basic assumptions of the model from QCD.

In the "Chashire Cat bag" approach \cite {Cat} the bag itself has
no physical significance. One just formally separates
the space-time in a bag region where QCD still acts and an exterior
region where QCD is integrated out by an effective action
formulated in hadronic degrees of freedom (for example
the Skyrme model), and a boundary. The location of the boundary
has no physical consequences and moreover fermionic degrees of freedom
can be replaced by bosonic ones.

The idea of describing elementary particles as strongly curved universes
was discussed by Frolov and Markov \cite{FM}, by Isham, Salam and
Strathdee \cite{ISS} and other authors, see for example \cite {SS}.
Black holes and  negatively
curved anti-de Sitter space could evidently  provide baglike structures.
Timelike geodesics in {\it AdS} are known periodic with frequency $1/R$
where $R$ is the {\it AdS} radius.
Therefore if one assumes that free quarks move along geodesics then one
gets
that they carry out harmonic oscillations within a sphere of radius $R$.

According to \cite{Mal} the large $N$ limit of supersymmetric
gauge theory is described by supergravity on {\it AdS} space. Similar
description was also proposed for large $ N$ gauge theories without
supersymmetry \cite{Wit2} including QCD. It is known that QCD in
the large $N$ limit is equivalent to a meson theory which is not
reduced to the non-linear sigma model or Skyrme model. It is a
complex unknown theory with an infinite number of elementary fields
\cite{WitN}.
Now by using the Maldacena conjecture one can try to describe this
unknown meson theory in term of the infinite number of Kaluza-Klein
modes. Probably in this way one gets the "Kaluza-Klein
corrections" to the sigma model and to the Skyrme model.
A gauge invariant version of the Skyrme model is considered in \cite{Fad}.

In the hybrid chiral bag model \cite{CT} a surface coupling with
an external pion field is introduced to restore chiral invariance
which was broken in the original MIT bag model.
The Lagrangian for the external pion field is that of the non-linear
$\sigma$-model or the Skirme model.  A specific solution
was used for the meson field where its isospin direction point radially,
the so called hedgehog solution.
The same solution arises in a model for baryons, originally proposed
by Skyrme and then revisited by Witten \cite{WitN}.
In this model the baryons are topologically stable solutions
in a non-linear $\sigma$-model describing  only meson fields.
The topological charge of the soliton  solution is identified with
the baryon number.

In the recent papers \cite{Wit2,RY,Mal2,Wit3,RTY,BISY}
the thermodynamics of black holes in {\it AdS} space was used for the
investigation of thermodynamics of gauge theories. An ansatz for
the Wilson loop in the large $N$ limit was proposed which involves
calculating the area of a probe string worldsheet in a supergravity
background.

 The ansatz for the Wilson loop
\be
W(C)=\frac{1}{N}<Tr P e^{\int_C A}>
\ee
used in \cite {RY, Mal2} is
\be
W(C)=e^{-I}
\ee
where
\be
I=\int d^2\sigma \sqrt{det h_{ab}},~~ h_{ab}=g_{MN}\partial_ax^M
\partial_bx^N
\ee
The action $I$ is the area
of the string worldsheet which at the boundary of {\it AdS} describes
the loop $C$.
The metric
$g_{MN}$ is a Schwarzschild-de-Sitter supergravity solution
describing the near horizon geometry of near extremal D3-branes
in type IIB string theory
\be
ds^2=\alpha '[\frac{U^2}{R^2}(-fdt^2+dx_i^2)+\frac{R^2}{fU^2}dU^2
+R^2d\Omega_5^2]
\ee
where $f=1-U_0^4/U^4, R^2=g_{YM}^2N$ and $U_0^4$ is proportional
to the energy density on the brane.

From this it seems natural to propose that the computation of the
volume of a probe membrane worldvolume in a supergravity background
gives an {\it AdS} bag model.  It naturally allows
the "Cheshire Cat" principle \cite {Cat}.
A  proposal for the expectation value of the bag operator
$B(M)$  \cite{Are} is
\be
<B(M)>=e^{-S}.
\ee
Here $M$ is a two-dimensional surface describing the boundary of the bag
(membrane) and $S$ is the volume of the membrane which at the
boundary of {\it AdS} describes the surface $M$,
\be
S=\int d^3 \sigma \sqrt{det h_{ab}},~~ h_{ab}=g_{MN}
\partial_ax^M\partial_bx^N
\ee
In this picture mesons in QCD correspond to black holes in gravity and
the $\sigma$-model or the Skyrme model should give an effective
description of black holes because they give a phenomenological
description of mesons.  A gravitational analogue of barions in
such an approach is not clear at the moment.
The boundary conditions (\ref{1}) and (\ref{2})
are set on the spacelike surface. The horizon of a black hole is a
light-like surface. Hence an analogue of (\ref{1}) and (\ref{2})
one has to set on a stretched horizon.

A  derivation of the phenomenological description of multiple
black holes one can get along the discussion in \cite{AV-ad}
where bunches of branes are considered.
The supergravity solution in type IIB string theory carrying
D3-brane charge has the form
\be
ds^2=f^{-1/2}(-dt^2+dx_i^2)+f^{1/2}dy^2_{\mu}
\ee
where $i=1,2,3,~\mu=4,5,6,7,8,9$ and $f=f(y)$ is a harmonic function.

If we consider two bunches of D3-branes located
at points $y^{(1)}$ and  $y^{(2)}$ then
\be
f= 1+4\pi g\alpha'[\frac{N_1}{|y-y^{(1)} |^4}
+\frac{N_2}{|y-y^{(2)} |^4}]
\ee
This metric describes two black holes.
There are  gravitational forces
between two bunches  and
it seems natural to think that this configuration
of branes is described by  super Yang-Mills theory
in the curved background.
From the previous discussion one suggests
that these black holes can be considered
as mesons and can be described by using the sigma model
or the Skyrme model lagrangians. The same reasoning can be  applied
to a configuration of an arbitrary number of non-extremal black
holes. The quantum Boltzmann statistics found in hadron physics
is relevant also to the physics of black holes, see \cite{Vol,AV-ad}
and refs therein.
In the case $gN_1>>gN_2>>1$ one has super
Yang-Mills theory in the 3-brane background.

I am   grateful to I. Aref'eva for useful comments.
The work  is supported  in part by  RFFI grant 96-01-00312
and INTAS grant 96-0698.

\end{document}